# A Theoretical Study of Process Dependence for Standard Two-Process Serial Models and Standard Two-Process Parallel Models

Ru Zhang, Yanjun Liu, and James T. Townsend, Indiana University

## Introduction

The question whether people can mentally process multiple pieces of information in parallel or in series has intrigued psychologists since the 19$^{th}$ century. Parallel and serial systems and models are very important special cases of the issue of mental architectures. We pause to observe that the "architecture" part is not meant to imply rigidity. In some situations, individual differences and differences across conditions can emerge (e.g., Townsend & Fific', 2004).

Many of the authors of this honorific volume, including the figure of our tribute for his long and distinguished career, Professor Hans-Georg Geissler, have contributed substantially to the literature on mental architecture and related issues. There are now hundreds of experimental and theoretical works bearing on these topics and even a brief survey is outside our present scope. However, the reader is encouraged to explore them more deeply in recent reviews (e.g., Algom, Eidels, Hawkins, Jefferson, & Townsend, 2015; Townsend, Yang, & Burns, 2011). Although much has been learned over the years about parallel and serial processes, it seems there is always more to understand.

Now, general models of either ilk, despite their seeming simplicity, can range over a terrain of rather immense complexity. Only mathematical specification of their parallel vs. serial nature must be invariant (see, e.g., Townsend & Ashby, 1983; Townsend & Wenger, 2004). Nonetheless, there are specific classes which seem to be canonical in their parallel or serial purity, so to speak. In fact, some authors have tended to confine their notions of serial or parallel nature to only these limited versions. We have never adhered to that philosophy but we do believe they warrant a special echelon in the taxonomy of mental architectures.

We refer to the canonical parallel class as *standard parallel models* and the canonical serial class as *standard serial models* (see, e.g., Algom, Eidels, Hawkins, Jefferson, & Townsend, 2015; Townsend & Wenger, 2004).

Immense efforts have been expended, both theoretically and experimentally, in the quest to test serial from parallel perceptual, cognitive and action systems apart from one another. It has turned out that despite the relative simplicity and diametrically opposed arrangement of serial vs. parallel architectures, experimental identification has proven to be quite challenging.

In fact, in a number of popular experimental designs intended to test serial models against parallel models, it can be shown that the models are mathematically equivalent (e.g., Townsend, 1972; Townsend & Ashby, 1983; Townsend & Wenger, 2004). This chapter does not touch on, or review any of this literature.

On the other hand, a number of experimental methods have been developed that are capable of assessing mental architectures at a deeper level (e.g., Schweickert, 1978; Schweickert & Townsend, 1989; Townsend, 1990; Townsend & Wenger, 2004). The present chapter is closer to this train of research in that it asks about fundamental properties that differentiate certain important sub-classes of parallel and serial systems. However, it must be recognized that any empirical methods growing out of this work will not be so general as some of those presented in the above papers.

Thus, instead of searching for models of the alternative class of models that perfectly or imperfectly mimic the targeted class (begun in Townsend, 1969, 1972, 1974, and completed in Townsend, 1976a; Townsend & Ashby, 1983), we seek to understand the diversity of certain central properties. Although our mathematical discoveries are specialized to $n = 2$ processes (items, channels, etc.), we put down our foundational definitions for arbitrary $n$.

All serial models assume that at every moment, only one process is being activated (Figure 1(a)). In the general case, the order of processing the *n* items can possess a probability distribution. So if *n* processes are considered, there are *n*! ways to arrange the order of them. The parallel models assume that every process starts simultaneously (Figure 1(b)) but they can terminate at different moments. Typically, any of the *n*! orders could also occur according to parallel processing models.

A special instance of ordering in a serial system would assign a single order to occur with probability 1. A parallel system cannot, due to its mathematical constraints, be equivalent to that special type of serial processing although it could mimic it to an arbitrarily degree of approximation (see, e.g., Townsend, 1972; Townsend & Ashby, 1983).

One central feature of a standard serial model is the axiom that the successive processing times are stochastically independent. A central assumption of standard parallel models is that their processing times are stochastically independent. However, the generic statistic associated with the serial processing times is called the *intercompletion time.* In general, parallel models, even the standard variety, will not predict stochastically independent intercompletion times though special cases can be made to do so (as in Townsend & Ashby, 1983, Chapter 4).

On the other hand, the generic statistic associated with the parallel processing time random variable is the *total completion time,* which can be recorded, of course, even when processing is serial. Serial models will not be expected to, in general, predict stochastically independent total completion times.

What standard parallel models predict in the way of intercompletion times, and what standard serial models predict for total completion times, are the subjects of our enquiry.

We now proceed to construct more quantitatively exact definitions. The traditional manner of constructing and describing mental architectures is through an event space (more formally, sigma space) based on the times of completion of items (used in a generic sense) across time (e.g., Townsend, 1972, 1974). However, deeper characterizations and differentiation of architectures can be found with event spaces that take into account finer grained state spaces associated with the items or processes. For instance, if items can be considered to be constituted by sets of features, then serial and parallel models become more diverse and, in specific experimental paradigms, less subject to model mimicking dilemmas (see, e.g., Townsend, 1976b, Townsend & Evans, 1983; Snodgrass & Townsend, 1980; Townsend & Ashby, 1983, Chapter 13). In the present study, we adhere to the traditional approach, but statements about parallel—serial mimicry will sometimes take the deeper approach into account.

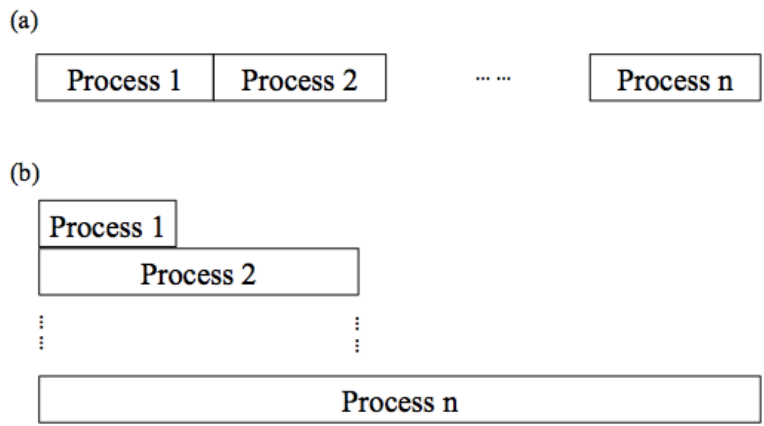

Figure 1. Examples of (a) a serial model and (b) a parallel model.

## 1. Terminology

Let us denote the processing time for process $j$ as $z_j$, where $j = 1, \dots, n$. In this paper, the term "processing time" is short for processing time of a process. The "$j$" in the term "process $j$" represents the identity of a process. We also use the term "the $j$-th process" in this paper to denote a process that is completed at $j$-th position. The "$j$" in the term "the $j$-th process" indicates that there are $j$-1 processes that have finished before that process. In Figure 1, process $j$ and the $j$-th process are the same process. One should keep in mind that they are usually different. In general, process $j$ may be the $i$-th process to complete, where $i \neq j$.

**Stage:** Stage $j$ spans the time interval from the completion of the ($j$-1)-th process to the completion of the $j$-th process. For serial models, each process corresponds to a stage. In Figure 1(a), process 1 is stage 1, process 2 is stage 2,…, and process $n$ is stage $n$. For parallel models, each stage covers a part of a process. In Figure 1(b), stage $j$ starts from the completion of process $j$-1 and terminates at the moment that process $j$ is complete.

**Intercompletion time:** The intercompletion time $T_j$ is the time that is spent for stage $j$. In this paper we also call $T_j$ the $j$-th intercompletion time. For the serial models, each processing time is an intercompletion time as anticipated in the preceding paragraph: In Figure 1(a), $T_1 = z_1, T_2 = z_2,…, T_n = z_n$. In contrast, for the parallel model in Figure 1(b), $T_1 = z_1, T_2 = z_2 - z_1,…, T_n = z_n - z_{n-1}$.

**Total completion time:** The total completion time $\mathbb{T}_j$ is the time that is consumed from the onset of the model to the moment that process $j$ is complete. For the serial model in Figure 1(a), $\mathbb{T}_1 = z_1$, $\mathbb{T}_2 = z_1 + z_2,…, \mathbb{T}_n = z_1 + z_2 + \cdots + z_n$. For parallel models, the total completion time is the processing time. In Figure 1(b), $\mathbb{T}_1 = z_1, \mathbb{T}_2 = z_2,…, \mathbb{T}_n = z_n$.

Note that the intercompletion time $T_j$ is defined with respect to the stage in the processing order, whereas the total processing time for a process $\mathbb{T}_j$ is defined with respect to the identity of a process.

## 2. Serial Models and Parallel Models

**Serial models:** One can represent the serial models using the product of the probability of a certain serial order of processes and the joint density function of intercompletion times conditioned on the order.
$$P(I)f_s(T_1 = t_1, …, T_n = t_n | I = (i_1, …, i_n)).$$
Here $t_1, …, t_n$ are realizations of $T_1, …, T_n$, $(i_1, …, i_n) \in \text{Perm}(n)$, where

$\mathrm{Perm}(n)$ is the set of all permutations of the naturals from 1 to $n$, and $P(I)$ is the probability of a particular permutation $I = (i_1, \ldots, i_n)$. For the serial models, the permutation $I$ means that the model starts with process $i_1$ and is connected by the onset of process $i_2$ after process $i_1$ is complete, and so on.

**Parallel models:** Parallel models can be written as the joint density function of total completion times of processes:
$$f_p(\mathbb{T}_1 = \tau_1, \ldots, \mathbb{T}_n = \tau_n; I),$$
where $\tau_1, \ldots, \tau_n$ are realizations of $\mathbb{T}_1, \ldots, \mathbb{T}_n$. For the parallel models, the permutation $I$ means that all the processes start simultaneously but process $i_1$ terminates first, process $i_2$ terminates second, and so on.

### 3. Assumptions

Scientists can make further restrictive but still quite general and reasonable assumptions about the serial models and parallel models to differentiate the two types of models. Perhaps the most widely used stipulation is that of *selective influence*. This notion was firstly proposed by Sternberg (1969) in his additive factors method. It states that manipulation of each factor only influences the process that is associated with that factor. *System factorial technology* (SFT) (Townsend & Nozawa, 1995) was developed to characterize different types to parallel and serial models based on that assumption. One can diagnose the mental arrangements according to the interaction contrast of survival functions of reaction time. This technology is further explored by other researchers (Schweickert, Giorgini, & Dzhafarov, 2000; Dzhafarov, Schweickert, & Sung, 2004; Yang, Fific', & Townsend, 2013; Zhang & Dzhafarov, 2015).

A limitation of SFT is that it has to be applied in the complete factorial design, in which each factor has two or more levels and the requirement of *stochastic dominance* has to be met as well. Not every experiment affords such a manipulation. Recent reviews mentioned earlier delve into such matters in more depth and also survey other methods of identifying architectures and related mechanisms (e.g., Algom, Eidels, Hawkins, Jefferson, & Townsend,

2015; Townsend, Yang, & Burns, 2011; Townsend & Wenger, 2004).

Another candidate assumption that can be imposed to the investigated system is within stage independence. This assumption states that the processes are independently executed within each stage. Since there is only one process in each stage in the serial model, this assumption is moot in the class of serial models. Although within stage independence is an important characteristic to know about, it turns out that within stage dependent models can be mathematically transformed to within stage independent models (e.g., see Rao, 1992, pp. 162-163). Thus, such models cannot be discriminated in the absence of observability of the within stage dependencies.

Next, one can consider the processing time independence assumption. That is, the processing time of a process is independent of another. For the serial models, processing time independence is equivalent to across stage independence since the actual processing times are equivalent to the intercompletion times. In contrast, for the parallel models, processing time of an item or channel is equivalent to the interval from the very beginning of processing (which we may recall, entails that every parallel process begins simultaneously), until that individual channel is finished. This is a statistic known generically as the *total completion time.* Independence of processing times, between say, 2 channels, is tantamount to independence of their total completion times.

### 4. Standard Serial Models and Standard Parallel Models

With the assumption of independently and identically distributed (iid) processing times, the serial models are named standard serial models and parallel models are named standard parallel models.

Suppose there are only two processes $a$ and $b$ in the models. The standard serial models are then named *standard two-process serial models* and the standard parallel models are named *standard two-process parallel models.*

Let us denote the processing times of processes $a$ and $b$ as $z_a$ and $z_b$ (recall that they are iid) and the density function as $f$. The corresponding distribution function is labeled as $F$. The survival function, the hazard function, and the cumulative hazard function are represented as

$$S = 1 - F,$$
$$h = \frac{f}{S},$$
$$H = \int_0^x h(t)\, dt = -\ln[S(x)].$$

**Standard two-process serial models:** Since two processes are under consideration, the model can be decomposed into two stages. If process $a$ is executed earlier than process $b$, then process $a$ is stage 1. If process $a$ is executed later than process $b$, then process $b$ is stage 1 (Figure 2).

Case I

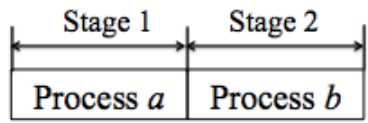

Case II

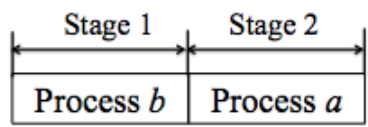

Figure 2. Possible process arrangements in a standard two-process serial model.

As defined earlier, the intercompletion time $T_1$ is the time that is spent for stage 1 and $T_2$ is the time that is spent for stage 2. So for Case I, $T_1 = z_a$, $T_2 = z_b$ and for Case II, $T_1 = z_b$, $T_2 = z_a$. It is apparent that $T_1$ and $T_2$ are iid as $z_a$ and $z_b$ are assumed iid. The total completion time $\mathbb{T}_a$ is the time that is consumed from the onset of the model to the moment that process $a$ is complete. The total completion time $\mathbb{T}_b$ is the time that is consumed from the onset of the model to the moment that process *b* is complete. So for Case I, $\mathbb{T}_a = T_1 = z_a$, $\mathbb{T}_b = T_1 + T_2 = z_a + z_b$ and for Case II, $\mathbb{T}_a = T_1 + T_2 = z_a + z_b$, $\mathbb{T}_b = T_1 = z_b$.

**Standard two-process parallel models:** Since two processes are under consideration (see an example in Figure 3), the total completion time for process *a* and the total completion time for process *b* are

$$\mathbb{T}_a = z_a,$$
$$\mathbb{T}_b = z_b,$$

respectively. Please note that Figure 3 is an exemplar representation of a standard two-process parallel model in which process $a$ is faster than process $b$. It is indeed possible that process $a$ is slower than process $b$ as $z_a$ and $z_b$ are iid. The intercompletion times in Figures 3 can be represented as

$$T_1 = \mathbb{T}_a = z_a,$$
$$T_2 = \mathbb{T}_b - \mathbb{T}_a = z_b - z_a.$$

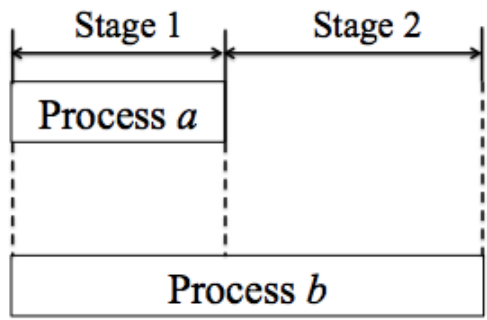

Figure 3. A standard two-process parallel model.

Townsend and Ashby (1983) showed that by assuming the distributions of processing times are exponential, standard two-process serial models yield a *positive dependence* between the total completion times of processes *a* and *b*. That is the conditional probability that *a* is completed before some time $\tau$ given *b* has already been completed by this time is greater than the unconditional probability that *a* is completed by time $\tau$. It has not been known if this prediction holds for total completion times with arbitrarily distributed processes.

Turning to the issue of dependence of intercompletion times, there are two intriguing questions that arise: 1. How does that second intercompletion time behave as the time occupied by the first termination (equal to the first intercompletion time) increases? Note that in a standard serial model, the second intercompletion time is independent of the first intercompletion time. 2. How does the second intercompletion time behave in comparison to the distribution (speed, etc.) of the first intercompletion time?

## 5. Experimental Paradigms Related to Intercompletion Times

Free recall is one of the experimental paradigms that was systematically used to study the serial-parallel issue. In this paradigm, the subject is instructed to recall words that belong to a semantic category from their long-term memory (Bousfield & Sedgewick, 1944; Bousfield, Sedgewick, & Cohen, 1954), for instance, name as many cities in the United States as they could remember. The words were reported successively. It was found that the time interval between two successive responses, that is the intercompletion time, increased as more responses were generated (Murdock & Okada, 1970; Patterson, Meltzer, & Mandler, 1971; Pollio, Kasschau, & DeNise, 1968; Pollio, Richards, & Lucas, 1969).

McGill contributed an influential chapter on stochastic processes in psychology, to the 1963 Volume 1 of the Handbook of Mathematical Psychology. His work is based on a serial model to account for the general temporal characteristics of the Bousfield & Sedgewick (1944) data. He assumed only one item could be sampled from a search set and inspected at any time. It assumes all the relevant items are equally likely to be chosen at each draw. After each draw, the subject inspects if the item is a member of the specified category and if the item is not recalled earlier before reporting this item. The amount of time for each inspection, which is the so called intercompletion time or processing time of each process (item) in the language of serial models, is assumed exponentially distributed with the same rate parameter. Interestingly, his serial model is identical to a standard parallel model with exponential distributions on the $n$ channels.

Since some items can be more easily to be accessed (Shiffrin, 1970), Vorberg and Ulrich (1987) generalized McGill's work to the models that assume unequal accessibility. The generalized model can remove the minor discrepancies between the original McGill's model and data in terms of predicting the number of generated items by a certain time moment. The stochastic representation of Vorberg and Ulrich's (1987) serial model is written as

$$P(I)f_s(T_1 = t_1, \ldots, T_n = t_n | I = (i_1, \ldots, i_n))$$
$$= \left(\prod_{j=1}^n \frac{u_{i_j}}{\sum_{l=j}^n u_{i_l}}\right) \left[\prod_{j=1}^n (\sum_{l=j}^n u_{i_l}) \exp(-\sum_{l=j}^n u_{i_l} t_j)\right],$$

where $u_{i_j}$ stands for the rate parameter for process (or item) $i_j$ and $n$ is the number of recallable target items within the search set. The model predicts that the rate parameter of the *j*-th intercompletion time equals the sum of rate parameters of processes that have not been executed. Note that when equal accessibility assumption is imposed ($u = u_{i_1} = \cdots = u_{i_n}$), the intercompletion time distribution does not depend on the recall order any more and the conditional joint density function is

$$f_s(T_1 = t_1, \ldots, T_n = t_n | I = (i_1, \ldots, i_n))$$
$$= \prod_{j=1}^n (n-j+1)u \exp[-(n-j+1)u t_j],$$

which is McGill (1963)'s model. Vorberg and Ulrich (1987) also derived the stochastic representation of the counterpart parallel model with the assumption of unequal accessibility:

$$f_p(\mathbb{T}_1 = \tau_1, \ldots, \mathbb{T}_n = \tau_n; I)$$
$$= \prod_{j=1}^n u_j \exp(-u_j \tau_j).$$

The total processing times are independent with each other in the counterpart parallel model.

According to McGill's model, Rohrer and Wixted (1994) derived a hyperbolic intercompletion time growth as more items are generated:

$$\bar{T}_j = \frac{1}{u(n-j)},$$

where $\bar{T}_j$ is the mean of the *j*-th intercompletion time. This equation reflects that an intercompletion time is uniquely determined $n - j$. Thus, the last intercompletion time of a four-item recall should equal the last intercompletion time of a nine-item recall. They conducted experiments by asking the subjects to recall the words studied earlier.

By manipulating the size of the study word list or/and the presentation time for each word, they found the hyperbolic intercompletion time growth fitted the data well. As mentioned earlier, McGill's serial model is identical to a standard parallel model with exponential processing times. Thus Rohrer and Wixted (1994) study is an example to support the iid processing time assumption if one considers the recall a parallel process.

## 6. Goals

Despite of all variations of serial and parallel models, in this paper we aim to (1) differentiate, characterize, and compare standard serial models with standard parallel models by investigating the behavior of (conditional) distributions, rather than the joint density function or the mean, of the total completion times and also the intercompletion times, in complete generality, that is, without assuming any particular form for the distributions of processing times, (2) investigate if either or both models can account for the growth of intercompletion time as a function of output position in free recall experiments, and (3) check if additional constraints should be imposed on the models that result in the theoretically derived behavior of the models consistent with the empirical finding.

For the current study, we confine our discussion on the standard two-process serial models and the standard two-process parallel models. Our work can likely be generalized to the models with $n > 2$ processes in an analogous way.

### Dependence of Total Completion Times

As stated above, our first goal is to investigate the dependence of total completion times without assuming a specific family distribution to processing times.

A natural way of doing this is to compare the distribution function of $\mathbb{T}_b$ conditional on $\mathbb{T}_a$ versus the unconditional distribution function of $\mathbb{T}_b$. That is

$$P(\mathbb{T}_b \leq \tau | \mathbb{T}_a \leq \tau) - P(\mathbb{T}_b \leq \tau). \quad (1)$$

If it is positive then we conclude that the total completion times in this case are positively dependent in a strong distributional sense, and conversely if the difference is negative. One can also investigate $P(\mathbb{T}_a \leq \tau | \mathbb{T}_b \leq \tau) - P(\mathbb{T}_a \leq \tau)$, which is indeed not different from (1).

## 1. Dependence of Total Completion Times in Standard Two-Process Serial Models

It was proven by Townsend & Ashby (1983, Page 73-74), if the processing times $z_a$ and $z_b$ (or the intercompletion times $T_1$ and $T_2$) in a two-process serial model are iid and follow exponential distributions, then $(1) > 0$ for $\tau > 0$. We now investigate if $(1) > 0$ holds for distributional free processing times.

**Theorem 1.** For a standard two-process serial model, $P(\mathbb{T}_b \leq \tau | \mathbb{T}_a \leq \tau) - P(\mathbb{T}_b \leq \tau)$ can be either negative or positive for $\tau > 0$.
*Proof.* First note that
$$P(\mathbb{T}_b \leq \tau | \mathbb{T}_a \leq \tau) = \frac{P(\mathbb{T}_b \leq \tau \cap \mathbb{T}_a \leq \tau)}{P(\mathbb{T}_a \leq \tau)}.$$
The numerator is a convolution of the density function and the distribution function:
$$\begin{aligned}
P(\mathbb{T}_b \leq \tau \cap \mathbb{T}_a \leq \tau) \\
= P(\max(\mathbb{T}_b, \mathbb{T}_a) \leq \tau) \\
= P(T_1 + T_2 \leq \tau) \\
= P(z_a + z_b \leq \tau) \\
= \int_0^\tau \int_0^y f(y-x)f(x)dxdy \\
= f(\tau) * F(\tau).
\end{aligned}$$
Note that it is always true that $f(\tau) * F(\tau) \leq 1$. The denominator is the probability that the process $a$ gets completed by time $\tau$ and is composed of the probability that $a$ is completed first (i.e., $p$) times the probability that it is completed by time $\tau$ (i.e., $F(\tau)$) plus the probability that $b$ is processed first (i.e., $1-p$) times the probability that both have been completed by that time (i.e., $f(\tau) * F(\tau)$). We then have

$$P(\mathbb{T}_a \leq \tau) = pF(\tau) + (1-p)f(\tau) * F(\tau).$$

On the other hand, the unconditional probability that $b$ gets finished by time $\tau$ is

$$P(\mathbb{T}_b \leq \tau) = (1-p)F(\tau) + pf(\tau) * F(\tau).$$

Consequently,

$$P(\mathbb{T}_b \leq \tau | \mathbb{T}_a \leq \tau) - P(\mathbb{T}_b \leq \tau)$$
$$= \frac{f(\tau)*F(\tau)}{pF(\tau)+(1-p)f(\tau)*F(\tau)} - [(1-p)F(\tau) + pf(\tau) * F(\tau)]$$
$$= R\left\{1 - F(\tau) - p(1-p)\left[(f(\tau) * F(\tau))^{\frac{1}{2}} - \frac{F(\tau)}{(f(\tau)*F(\tau))^{\frac{1}{2}}}\right]^2\right\}, \quad (2)$$

where

$$R = \frac{f(\tau) * F(\tau)}{pF(\tau) + (1-p)f(\tau) * F(\tau)} \geq 0,$$

for $\tau > 0$. Since $p(1-p) \leq \frac{1}{4}$, then

$$(2) \geq R\left\{1 - F(\tau) - \frac{1}{4}\left[(f(\tau) * F(\tau))^{\frac{1}{2}} - \frac{F(\tau)}{(f(\tau)*F(\tau))^{\frac{1}{2}}}\right]^2\right\}.$$

Note that if $p = \frac{1}{2}$, the above $\geq$ reduces to $=$. Now we need to investigate the sign of this term:

$$\left\{1 - F(\tau) - \frac{1}{4}\left[(f(\tau) * F(\tau))^{\frac{1}{2}} - \frac{F(\tau)}{(f(\tau)*F(\tau))^{\frac{1}{2}}}\right]^2\right\}. \quad (3)$$

If (3) >0, it indicates that (1) is positive. In other words, positive dependence holds in standard two-process serial models with distributional free processing times. Otherwise positive dependence does not hold. The sign of (3) was estimated using the computational method. We have

$$f(\tau) * F(\tau) = P(z_a + z_b \leq \tau) = P(z_a \leq \tau - z_b),$$
$$F(\tau) = P(z_a \leq \tau).$$

Therefore $f(\tau) * F(\tau) \leq F(\tau)$ since $z_b \geq 0$. Having $0 \leq f(\tau) * F(\tau) \leq F(\tau) \leq 1$ for $\tau > 0$,

$$\frac{F(\tau)}{(f(\tau) * F(\tau))^{\frac{1}{2}}} = \left[\frac{F(\tau)F(\tau)}{f(\tau) * F(\tau)}\right]^{\frac{1}{2}}$$

$$= \left[\frac{P(z_a \leq \tau)P(z_b \leq \tau)}{f(\tau) * F(\tau)}\right]^{\frac{1}{2}}$$

$$= \left[\frac{P(\max(z_a, z_b) \leq \tau)}{f(\tau) * F(\tau)}\right]^{\frac{1}{2}} \geq 1.$$

The computation was conducted with the order constraints stated above. In addition, we let $f(\tau) * F(\tau)$ vary uniformly within the interval $[0,1]$ and $F(\tau)$ vary uniformly within the interval $[f(\tau) * F(\tau), 1]$. The computational steps are:

Step 1: generate a random number $\alpha \sim \text{Uniform}[0,1]$, where $\alpha$ represents $f(\tau) * F(\tau)$.

Step 2: generate a random number $\beta \sim \text{Uniform}[\alpha, 1]$, where $\beta$ represents $F(\tau)$.

Step 3: if $\frac{\beta^2}{\alpha} \geq 1$, then compute (3) to check if it is strictly larger than 0.

1,000,000 pairs of $(\alpha, \beta)$ were generated. It was found that the probability of (3) $> 0$ conditional on $\frac{\beta^2}{\alpha} \geq 1$ was 62%. The computational result indicates that the sign of $P(\mathbb{T}_b \leq \tau | \mathbb{T}_a \leq \tau) - P(\mathbb{T}_b \leq \tau)$ is not definite. $\square$

According to the computation above, one may conclude that positive dependence for the total completion times does not necessarily hold in the standard two-process serial models if no specific distributions are imposed to processing times. Here we construct examples to provide more illustrations about it. In the examples, we assume the processing times $z_a$ and $z_b$ (or equivalently $T_1$ and $T_2$) follow Weibull distributions, exponential distributions, which is a special case of Weibull distributions, or uniform distributions. Uniform distributions are not usually used to model time variables. We discuss this type of distribution as it results in the interesting behavior of process dependence. We observe positive dependence holds when the processing times are Weibull distributed with the parameter $k = .5, 1$ (that is the exponential distribution), and $1.5$ but fails when $k = .2, 2$, or the processing times are uniformly distributed.

**Weibull distributions**: Let $z_a, z_b$ be iid and follow the Weibull distribution with the density function
$$f(\tau) = ku(u\tau)^{k-1}\exp[-(u\tau)^k],$$
where the parameters $k, u > 0$. Since the convolution of two Weibull variables does not have an analytic form, we use the computational method to achieve the values for (3). We present the 3d plots for (3) by varying the values of $\tau$ and $u$ (Figures 4 and 5). We allow $u$ to vary from .5 to 10 and $\tau$ to vary from 0.01 to 5. In Figure 4, the upper plot fixes $k = .5$ and the bottom one fixes $k = 1.5$. The values of (3), represented by the vertical coordinate, are positive and approaching zero as $\tau \to \infty$. In Figure 5, the upper plot fixes $k = .2$ and the bottom one fixes $k = 2$. The values of (3), are non positive when $k = .2$ and fluctuate from positive to negative and then approach zero as $\tau \to \infty$ in general when $k = 2$.

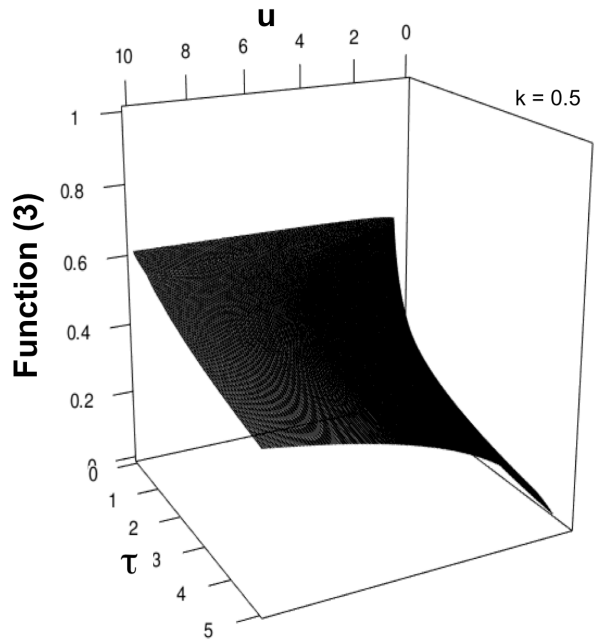

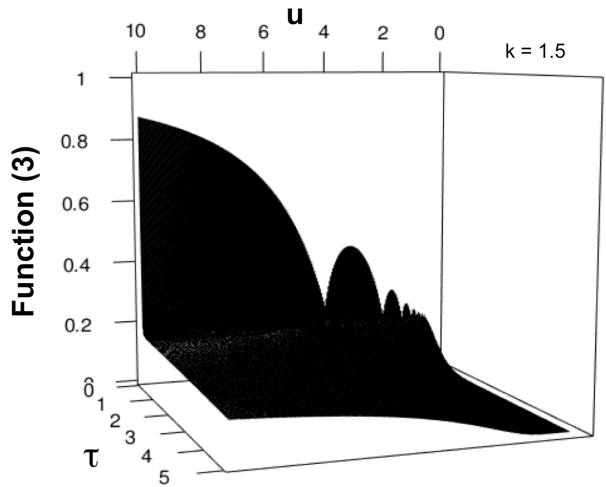

Figure 4. Plots of (3) for Weibull distributed processing times, where $k = .5$ (upper) and $k = 1.5$ (bottom).

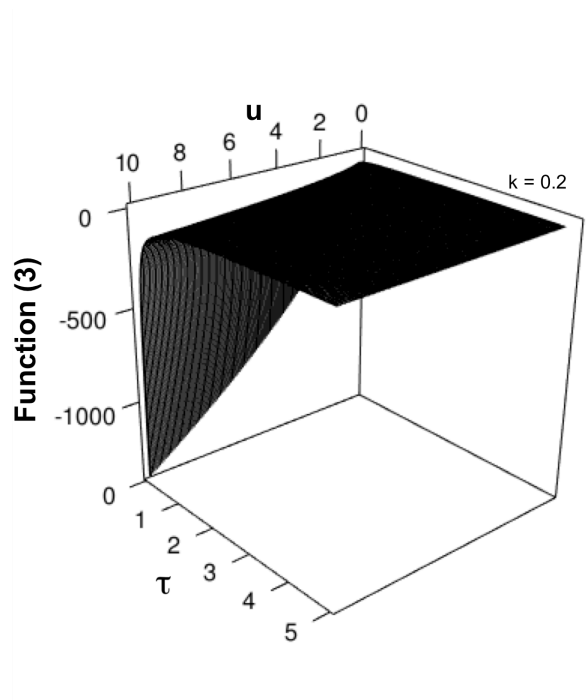
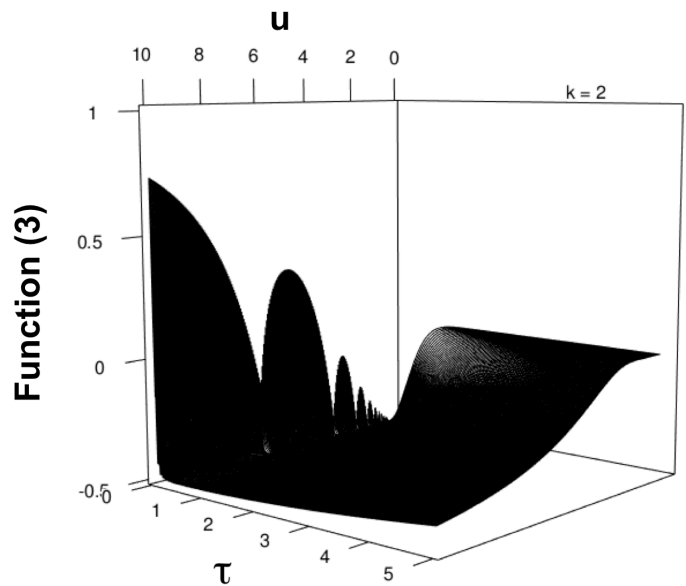

Figure 5. Plots of (3) for Weibull distributed processing times, where $k = .2$ (upper) and $k = 2$ (bottom).

**Exponential distributions** (also see Townsend & Ashby, 1983, Page 73-74): When $k = 1$, Weibull distributions reduce to exponential distributions. Now the density function for $z_a, z_b$ is
$$f(\tau) = u\exp(-u\tau).$$
The corresponding distribution function is
$$F(\tau) = 1 - e^{-u\tau}.$$
Also we have
$$f(\tau) * F(\tau) = \int_0^\tau ue^{-u(\tau-\tau_2)}(1 - e^{-u\tau_2})d\tau_2 = 1 - e^{-u\tau} - u\tau e^{-u\tau}.$$
We then have
$$P(\mathbb{T}_b \leq \tau | \mathbb{T}_a \leq \tau) - P(\mathbb{T}_b \leq \tau)$$
$$= \frac{1}{pF(\tau) + (1-p)f(\tau) * F(\tau)}\{f(\tau) * F(\tau)$$
$$\quad - [pF(\tau) + (1-p)f(\tau) * F(\tau)][(1-p)F(\tau) + pf(\tau)$$
$$\quad * F(\tau)]\}$$
$$= R'\{1 - e^{-u\tau} - u\tau e^{-u\tau}$$
$$\quad - [p(1 - e^{-u\tau}) + (1-p)(1 - e^{-u\tau} - u\tau e^{-u\tau})][(1-p)(1$$
$$\quad - e^{-u\tau}) + p(1 - e^{-u\tau} - u\tau e^{-u\tau})]\}$$
$$= R'\{1 - e^{-u\tau} - u\tau e^{-u\tau}$$
$$\quad - [1 - 2e^{-u\tau} - u\tau e^{-u\tau} + e^{-2u\tau} + u\tau e^{-2u\tau} + p(1$$
$$\quad - p)u^2\tau^2 e^{-2u\tau}]\}$$
$$= R' e^{-u\tau}[1 - e^{-u\tau} - u\tau e^{-u\tau} - p(1-p)u^2\tau^2 e^{-u\tau}],$$
where $R' = \frac{1}{pF(\tau)+(1-p)f(\tau)*F(\tau)}$. Since $p(1-p) \leq \frac{1}{4}$, the above expression is
$$\geq R' e^{-u\tau}\left[1 - e^{-u\tau} - u\tau e^{-u\tau} - \frac{1}{2}u^2\tau^2 e^{-u\tau}\right]$$
$$> 0$$
for $\tau > 0$ as the distribution function of a three-stage gamma with rate $u$ is just $1 - e^{-u\tau} - u\tau e^{-u\tau} - \frac{1}{2}u^2\tau^2 e^{-u\tau}$.

**Uniform distributions**: Let $z_a, z_b$ be iid and follow the uniform distribution:
$$z_a, z_b \sim \text{Uniform}(0, v),$$
where $v > 0$. The corresponding distribution function is
$$F(\tau) = \begin{cases} \frac{\tau}{v}, & \text{if } 0 \leq \tau < v \\ 1, & \text{otherwise} \end{cases}.$$
Also we have

$$f(\tau) * F(\tau) = \begin{cases} \int_0^\tau \frac{\tau}{v^2} d\tau_2 = \frac{\tau^2}{2v^2}, \text{if } 0 \leq \tau < v, \\ \int_v^\tau \frac{2v - \tau_2}{v^2} d\tau_2 + \frac{1}{2} = \frac{2\tau}{v} - \frac{\tau^2}{2v^2} - 1, \text{if } v \leq \tau < 2v, \\ 1, \text{if } 2v \leq \tau. \end{cases}$$

We then have for $2v \leq \tau$,
$$P(\mathbb{T}_b \leq \tau | \mathbb{T}_a \leq \tau) - P(\mathbb{T}_b \leq \tau) = 0;$$
for $v \leq \tau < 2v$,
$$P(\mathbb{T}_b \leq \tau | \mathbb{T}_a \leq \tau) - P(\mathbb{T}_b \leq \tau)$$
$$= R \left\{ 1 - F(\tau) - p(1-p) \left[ (f(\tau) * F(\tau))^{\frac{1}{2}} - \frac{F(\tau)}{(f(\tau) * F(\tau))^{\frac{1}{2}}} \right]^2 \right\}$$
$$= R \left\{ 1 - 1 - p(1-p) \left[ (f(\tau) * F(\tau))^{\frac{1}{2}} - \frac{1}{(f(\tau) * F(\tau))^{\frac{1}{2}}} \right]^2 \right\}$$
$$\leq 0;$$
for $0 \leq \tau < v$,

$$P(\mathbb{T}_b \leq \tau | \mathbb{T}_a \leq \tau) - P(\mathbb{T}_b \leq \tau)$$
$$= R \left\{ 1 - F(\tau) - p(1-p) \left[ (f(\tau) * F(\tau))^{\frac{1}{2}} - \frac{F(\tau)}{(f(\tau) * F(\tau))^{\frac{1}{2}}} \right]^2 \right\}$$
$$= R \left\{ 1 - \frac{\tau}{v} - p(1-p) \left[ \left(\frac{\tau^2}{2v^2}\right)^{\frac{1}{2}} - \frac{\frac{\tau}{v}}{\left(\frac{\tau^2}{2v^2}\right)^{\frac{1}{2}}} \right]^2 \right\}.$$

When $p = \frac{1}{2}$, the above expression is
$$= R \left[ 1 - \frac{\tau}{v} - \frac{1}{4} \left( \frac{\sqrt{2}\tau}{2v} - \sqrt{2} \right)^2 \right]$$
$$= R \left[ 1 - \frac{1}{2} \left( \frac{\tau}{2v} + 1 \right)^2 \right].$$

It can be negative when e.g., $\frac{\tau}{v} = \frac{5}{6}$ or positive when e.g., $\frac{\tau}{v} = \frac{1}{2}$.

Theorem 1 considers the two possible permutations of process $a$ and process $b$. It is found that the dependence of total completion times can be both positive and negative, that is (1) can be $> 0, = 0$, or $< 0$ at different $\tau$. Corollary 2 states if only one permutation is allowed in the investigated system, dependence of total completion times is non negative, that is $(1) \geq 0$. Lemma 3 presents the covariance of a total completion times.

**Corollary 2.** For a standard two-process serial model, $P(\mathbb{T}_b \leq \tau | \mathbb{T}_a \leq \tau) - P(\mathbb{T}_b \leq \tau)$ is non-negative for $\tau > 0$ if only the Case I arrangement or only the Case II arrangement is allowed in the investigated system.
*Proof.* Under this situation, $p$ is either 0 or 1. So we have
$$(2) = R\left[1 - F(\tau)\right] \geq 0$$
for $\tau > 0$. □

**Lemma 3.** For a standard two-process serial model, if only the Case I arrangement or only the Case II arrangement is allowed, $\text{Cov}(\mathbb{T}_a, \mathbb{T}_b) = \text{Var}(T_1)$.
*Proof.* $\text{Cov}(\mathbb{T}_a, \mathbb{T}_b) = \text{E}[(T_1)(T_1 + T_2)] - \text{E}(T_1)\text{E}(T_1 + T_2)$
$= \text{E}(T_1^2) + \text{E}(T_1 T_2) - \text{E}^2(T_1) - \text{E}(T_1)\text{E}(T_2)$
$= \text{Var}(T_1).$ □

## 2. Dependence of Total Completion Times in Standard Two-Process Parallel Models

Similarly as in standard two-process serial models, we investigate the behavior of (1) in standard two-process parallel models without assuming any specific distributions to processing times $z_a$ and $z_b$. It is found that $(1) = 0$ and this statement is presented in Theorem 4. We also compute the covariance of a total completion times for standard two-process parallel models. The result is presented in Lemma 5.

**Theorem 4.** For a standard two-process parallel model, $P(\mathbb{T}_b \leq \tau | \mathbb{T}_a \leq \tau) - P(\mathbb{T}_b \leq \tau) = 0$ for $\tau > 0$.
*Proof.* Recall that for a standard two-process parallel model, $\mathbb{T}_a$ and $\mathbb{T}_b$ are iid since $z_a$ and $z_b$ are iid. So we have

$$P(\mathbb{T}_b \leq \tau | \mathbb{T}_a \leq \tau) - P(\mathbb{T}_b \leq \tau)$$
$$= \frac{P(\mathbb{T}_b \leq \tau \cap \mathbb{T}_a \leq \tau)}{P(\mathbb{T}_a \leq \tau)} - P(\mathbb{T}_b \leq \tau)$$
$$= \frac{P(\mathbb{T}_b \leq \tau)P(\mathbb{T}_a \leq \tau)}{P(\mathbb{T}_a \leq \tau)} - P(\mathbb{T}_b \leq \tau)$$
$$= 0. \qquad \square$$

**Lemma 5.** For a standard two-process parallel model, $\text{Cov}(\mathbb{T}_a, \mathbb{T}_b) = 0$.

*Proof.* This is apparent. $\square$

Standard two-process serial models and standard two-process parallel models can be differentiated according to Theorem 1 and Theorem 4. Specifically, $P(\mathbb{T}_b \leq \tau | \mathbb{T}_a \leq \tau) - P(\mathbb{T}_b \leq \tau)$ cannot always be zero for a standard two-process serial model; while as for a standard two-process parallel model, the function maintains zero along the axis of $\tau$. One can also differentiate the two models according to Lemma 3 and Lemma 5.

## Dependence of Intercompletion Times

The intercompletion times in a standard serial model are processing times that are assumed independent and identically distributed. Therefore the empirical finding that as the number of stage increases, the intercompletion time also increases cannot be accounted by the standard serial models. It does not imply investigating the theories of standard serial models valueless. If one studies for instance typewriting, it is very possible that the intercompletion times for typing letters are independently and identically distributed. Then the standard serial model is applicable in this paradigm.

In contrast, standard parallel models can account for this phenomenon. Without loss of generalization, we label the process completed earlier process *a* and the other is labeled as process *b* in a standard two-process parallel model. Recall that the processing times $z_a$ and $z_b$, or equivalently $\mathbb{T}_a$ and $\mathbb{T}_b$, are assumed iid. Now let us label the intercompletion times for stage 1 and stage 2 as $T_a$ and $T_b$, where

$$T_a = \mathbb{T}_a = z_a,$$
$$T_b = \mathbb{T}_b - \mathbb{T}_a = z_b - z_a.$$

We investigate the survival function of the inter completion time $T_b$ conditional on the completion of stage 1: $P(T_b > t | \mathbb{T}_b > \mathbb{T}_a)$, where $t > 0$. Interestingly, it is found that the behavior of $P(T_b > t | \mathbb{T}_b > \mathbb{T}_a)$ depends on the hazard function of processing time $h$.

**Lemma 6.** For a standard two-process parallel model, if the hazard function $h$ is non-increasing, then $P(T_b > t | \mathbb{T}_b > \mathbb{T}_a)$ is non-decreasing as $T_a$ is increased.

*Proof.* We have
$$P(T_b > t | \mathbb{T}_b > \mathbb{T}_a)$$
$$= P(T_a + T_b > T_a + t | \mathbb{T}_b > \mathbb{T}_a)$$
$$= P(\mathbb{T}_b > T_a + t | \mathbb{T}_b > T_a)$$
$$= \frac{S(T_a + t)}{S(T_a)},$$
which is a ratio of two survival functions. To examine the behavior of this ratio as $T_a$ changes, one can take the derivative:
$$\frac{d}{dT_a} \frac{S(T_a + t)}{S(T_a)}$$
$$= \frac{-S(T_a)f(T_a + t) + S(T_a + t)f(T_a)}{S^2(T_a)}$$
$$= \frac{S(T_a)S(T_a + t)}{S^2(T_a)} \left[ \frac{f(T_a)}{S(T_a)} - \frac{f(T_a + t)}{S(T_a + t)} \right]$$
$$= \frac{S(T_a)S(T_a + t)}{S^2(T_a)} [h(T_a) - h(T_a + t)].$$

If the hazard function $h$ is non-increasing, then $\frac{d}{dT_a} \frac{S(T_a+t)}{S(T_a)} \geq 0$. Consequently, $P(T_b > t | \mathbb{T}_b > \mathbb{T}_a)$ is non-decreasing as $T_a$ is increased. □

A separate issue is how does the later stage (stage 2) compete with the earlier stage (stage 1) in a standard two-process parallel model. Let us denote the ratio of the hazard functions:
$$\alpha(t, T_a + t) = \frac{h(T_a + t)}{h(t)}.$$

The survival function at stage 1 in the standard two-process parallel model is $S^2(t)$. The survival function at stage 2 is $P(T_b > t | \mathbb{T}_b > \mathbb{T}_a)$, which is equivalent to $\frac{S(T_a+t)}{S(T_a)}$. We then aim to investigate $S^2(t)$ vs. $\frac{S(T_a+t)}{S(T_a)}$. If the survival from stage 1 to stage 2 is increasing, this trend is then consistent with the empirical finding that the amount of intercompletion time grows as the number of stages grows. Theorem 7 provides under what exact condition the survival increases from stage 1 to 2. Corollary 8 states that standard two-process parallel models with concave or linear cumulative hazard function $H(t)$ result in the increasing survival from stage 1 to stage 2.

**Theorem 7.** The survival function from the first stage to the second stage in a standard two-process parallel model is non-increasing or increasing depending on the value of $\alpha(t, T_a + t)$. That is $S^2(t) - \frac{S(T_a+t)}{S(T_a)} \geq 0$ if $\alpha(t, T_a + t) \geq 2$; $S^2(t) - \frac{S(T_a+t)}{S(T_a)} < 0$ if $\alpha(t, T_a + t) < 2$.

*Proof.* Note that $S(t) = \exp[-H(t)]$. We have
$$S^2(t) - \frac{S(T_a + t)}{S(T_a)}$$
$$= \exp[-2H(t)] - \frac{\exp[-H(T_a + t)]}{\exp[-H(T_a)]}.$$

The sign of the above equation is the same as the function below

$$-2H(t) + H(T_a + t) - H(T_a), \qquad (4)$$

which is equivalent to
$$\int_0^t -2h(t)dt + \int_{T_a}^{T_a+t} h(t)dt$$
$$= \int_0^t -2h(t)dt + \int_0^t h(T_a + t)dt$$
$$= \int_0^t [-2h(t) + \alpha(t, T_a + t)h(t)]dt$$
$$\begin{cases} \geq 0, \text{ if } \alpha(t, T_a + t) \geq 2 \\ < 0, \text{ if } \alpha(t, T_a + t) < 2 \end{cases}.$$

Therefore $S^2(t) - \frac{S(T_a+t)}{S(T_a)} \geq 0$ if $\alpha(t, T_a + t) \geq 2$ and otherwise $< 0$.
□

**Corollary 8.** For a standard two-process parallel model, (i) if the cumulative hazard function $H(t)$ is concave or linear, then $S^2(t) - \frac{S(T_a+t)}{S(T_a)} < 0$; (ii) if $H(t)$ is convex, then the sign of $S^2(t) - \frac{S(T_a+t)}{S(T_a)}$ is uncertain.

*Proof.* (i) If the cumulative hazard function $H(t)$ is concave or linear, then the hazard function $h(t)$ is decreasing or a constant. Consequently, $\alpha(t, T_a + t) < 1$ or $= 1$. According to Theorem 7, $S^2(t) - \frac{S(T_a+t)}{S(T_a)} < 0$ is proved. (ii) If $H(t)$ is convex, the hazard function $h(t)$ is increasing. Then we have $\alpha(t, T_a + t) = \frac{h(T_a+t)}{h(t)} > 1$. It is uncertain if $\alpha(t, T_a + t) \geq 2$ or $\alpha(t, T_a + t) < 2$. So the sign of $S^2(t) - \frac{S(T_a+t)}{S(T_a)}$ is uncertain. □

Similar as in the section of dependence of total completion times for standard two-process serial models, we construct examples to further illustrate the behavior of $S^2(t) - \frac{S(T_a+t)}{S(T_a)}$ for standard two-process parallel models. In the examples, we assume the processing times $z_a$ and $z_b$ (or equivalently $\mathbb{T}_a$ and $\mathbb{T}_b$) follow Weibull distributions (exponential distributions are included as a special case) or uniform distributions. We observe when $k \leq 1$, the Weibull distributed $z_a$ and $z_b$ result in the increasing survival function from the first stage to the second stage. The survival function is neither increasing nor decreasing when $z_a$ and $z_b$ are Weibull distributed with e.g. $k = 2$ and $4$, or uniformly distributed.

**Weibull distributions:** We assume
$$z_a, z_b \sim \text{Weibull}(k, u),$$
where $k, u > 0$. The corresponding cumulative hazard function and the hazard function are
$$H(t) = u(ut)^{k-1}t$$
and
$$h(t) = uk(ut)^{k-1}.$$
If $k = 1$, then the Weibull distribution reduces to the exponential distribution:
$$z_a, z_b \sim \text{Exp}(u).$$

The cumulative hazard function for the exponential distribution is linear:
$$H(t) = ut.$$
The hazard function is a constant:
$$h(t) = u.$$
It is apparent
$$\alpha(t, T_a + t) = \frac{h(T_a+t)}{h(t)} = 1 < 2.$$
Therefore the survival function for exponentially distributed processes is increasing from the first stage to the second stage.

If $k < 1$, then the cumulative hazard function is concave and the hazard function is decreasing. Hence the survival is also increasing from stage 1 to stage 2 as
$$\alpha(t, T_a + t) = \frac{h(T_a+t)}{h(t)} < 1.$$
If $k > 1$, then the cumulative hazard function is convex as $\alpha(t, T_a + t) = \frac{h(T_a+t)}{h(t)} > 1$. It is uncertain if $\alpha(t, T_a + t) \geq 2$ or $\alpha(t, T_a + t) < 2$.
We investigate the sign of $S^2(t) - \frac{S(T_a+t)}{S(T_a)}$ by computing (4) using the computational method. We present the 3d plots (Figure 6) for (4) by varying the values of $t$ and $T_a$ from 0 to 10 and fixing $u = 1$. The upper plot fixes $k = 2$ and the bottom one fixes $k = 4$. We observe (4) can be negative and positive for different combinations of $t$ and $T_a$. We conclude that for $k > 1$, the survival function from stage 1 to stage 2 can be neither increasing nor decreasing.

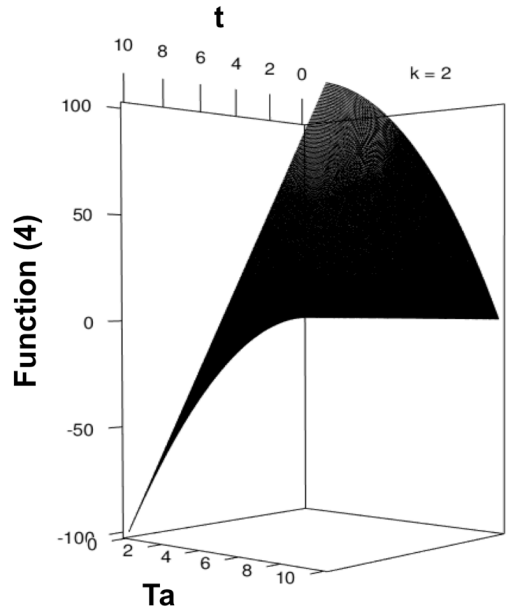

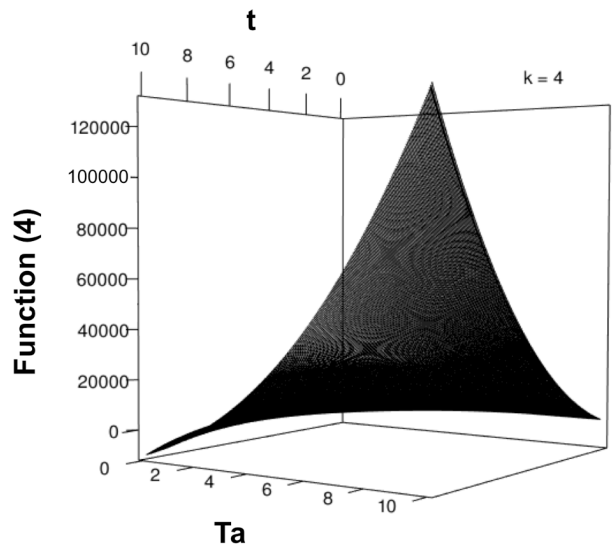

Figure 6. Plots of (4) for Weibull distributed processing times, where $k = 2$ (upper) and $k = 4$ (bottom).

**Uniform distributions**: Let
$$z_a, z_b \sim \text{Uniform}(0, v),$$
where $v > 0$. The corresponding cumulative hazard function is convex:
$$H(t) = -\ln(v - t) + \ln v$$
and the hazard function is
$$h(t) = \frac{1}{v - t}.$$
We have
$$\alpha(t, T_a + t) = \frac{h(T_a + t)}{h(t)} = \frac{v - t}{v - T_a - t} \geq 1.$$
We investigate the sign of $S^2(t) - \frac{S(T_a + t)}{S(T_a)}$ by computing (4) using the computational method. We present the 3d plot (Figure 7) for (4) by varying the values of $t$ and $T_a$ from 0 to 1 and fixing $v = 2$. We observe (4) can be negative and positive for different combinations of $t$ and $T_a$. Therefore the survival function from stage 1 to stage 2 is neither increasing nor decreasing for uniformly distributed processing times.

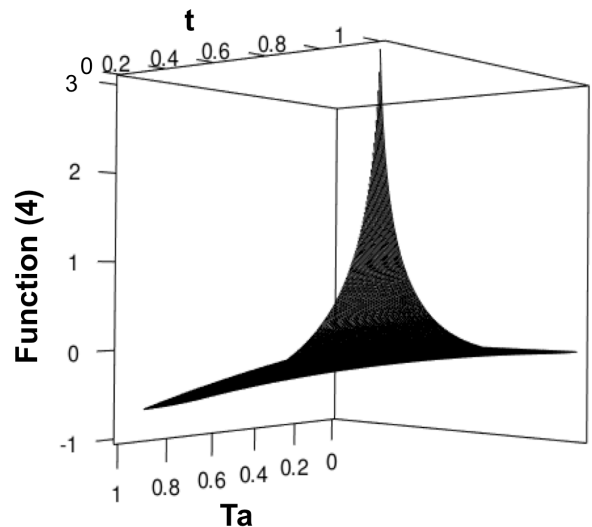

Figure 7. A plot of (4) for uniformly distributed processing times.

# Summary and Conclusions

In this article we differentiate and characterize the standard two-process serial models and the standard two-process parallel models by investigating the behavior of (conditional) distributions of the total completion times and survivals of intercompletion times without assuming any particular forms for the distributions of processing times. We address our argument through mathematical proofs and computational methods.

It is found that for the standard two-process serial models, positive dependence between the total completion times does not hold if no specific distributional forms are imposed to the processing times. That is the conditional probability that *a* is completed before some time $\tau$ given *b* has already been completed by this time can be greater or less than the unconditional probability that *a* is completed by time $\tau$. By contrast, for the standard two-process parallel models the total completion times are independent in the sense that the conditional probability that *a* is completed before some time $\tau$ given *b* has already been completed by this time is equal to the unconditional probability that *a* is completed by time $\tau$. According to different nature of process dependence, one can distinguish a standard two-process serial model from a standard two-process parallel model.

We also find that in standard two-process parallel models the monotonicity of survival function of the intercompletion time of stage 2 conditional on the completion of stage 1 depends on the monotonicity of the hazard function of processing time. We also find that the survival of intercompletion time(s) from stage 1 to stage 2 is increasing when the ratio of hazard function meets certain criterion. Then the empirical finding that the intercompletion time is grown with the growth of the number of recalled words can be accounted by standard parallel models. We also find that if the cumulative hazard function is concave or linear, the survival from stage 1 to stage 2 is increasing.

## 1. Limitations

We understand that the iid assumption for processing times is strong as in many paradigms it may not hold. Nevertheless this assumption may hold in some specific paradigms, for instance in Rohrer and Wixted (1994). Moreover, with the iid assumption, the standard serial models and standard parallel models can be diagnosed according to the mathematical properties of the temporal variables. The standard parallel models take account of the positive association between the amount of intercompletion time and the number of items that have been generated. Therefore, the standard models are worthy a full theoretical investigation for the empirical and mathematical interest.

## 2. A Simple Recall Experiment to Illustrate the Models

We consider an experimental paradigm in which the subjects recall words from a previously learned list. Positive association between the amount of intercompletion time and number of recalled words is expected.

Standard serial models cannot interpret this result as we have discussed earlier. For the recall experiment that we propose currently, standard parallel models can be used to account for the data. Recall that the total completion times are iid in a standard parallel model. The experimenter records the total completion time of each recalled word, which are denoted as $\mathbb{T}_1, \cdots, \mathbb{T}_n$. Let us assume each total completion time follows the Weibull distribution:
$$f(\mathbb{T}_j) = ku(u\mathbb{T}_j)^{k-1}\exp\left[-(u\mathbb{T}_j)^k\right],$$
where $j \in \{1, \cdots, n\}$. The likelihood function can be written as
$$L = f(\mathbb{T}_1)f(\mathbb{T}_2)\cdots f(\mathbb{T}_n).$$
One can use maximum likelihood method to estimate the parameters $k$ and $u$ for the Weibull distribution. We expect that the estimated value of $k$ is less than 1, which is consistent with the prediction of Theorem 7 and Corollary 8.

# References

Algom, D. Eidels, A., Hawkins, R. X. D., Jefferson, B. & Townsend, J. T. (2015). Features of response times: Identification of cognitive mechanisms through mathematical modeling. Chapter in J. Busemeyer, Wang, J., Eidels, A. and Townsend, J. T. (Eds.). Handbook of Computational and Mathematical Psychology.

Bousfield, W. A., & Sedgewick, C. H. W. (1944). An analysis of sequences of restricted associative responses. Journal of General Psychology, 30, 149-165.

Bousfield, W. A., Sedgewick, C. H. W., & Cohen, B. H. (1954). Certain temporal characteristics of the recall of verbal associates. American Journal of Psychology, 67, 11 l-l 18.

Dzhafarov, E.N., Schweickert, R., & Sung, K. (2004). Mental architectures with selectively influenced but stochastically interdependent components. Journal of Mathematical Psychology, 48, 51-64.

McGill. W. J. (1963). Stochastic latency mechanisms. In R. D. Lute, R. B. Bush, & E. Galanter (Eds.), Handbook of mathematical psychology (Vol. 1). New York: Wiley.

Murdock, B. B., Jr., & Okada, R. (1970). Interresponse times in single-trial free recall. Journal of Experimental Psychology, 86, 263-267.

Patterson, K. E., Meltzer, R. H., & Mandler, G. (1971). Interresponse times in categorized free recall. Journal of Verbal Learning & Verbal Behavior, 10, 417-426.

Pollio, H. R., Kasschau, R. A., & DeNise, H. E. (1968). Associative structure and the temporal characteristics of free recall. Journal of Experimental Psychology, 76, 190-197.

Pollio, H. R., Richards, S., & Lucas, R. (1969). Temporal properties

of category recall. Journal of Verbal Learning & Verbal Behavior, 8, 529-536.

Rao, B.L.S. Prakasa.  (1992). Identifiability in stochastic models:  Characterization of probability distributions.  New York: Academic Press.

Rohrer, D., & Wixted, J. T. (1994). An analysis of latency and interresponse time in free recall. 22(5), 511-524.

Schweickert, R. (1978).  A critical path generalization of the additive factor method: Analysis of a Stroop task.  Journal of Mathematical Psychology, 18, 105-139.

Schweickert, R., Giorgini, M., & Dzhafarov, E.N. (2000). Selective influence and response time cumulative distribution functions in serial-parallel task networks. Journal of Mathematical Psychology, 44, 504-535.

Schweickert, R., & Townsend, J. T. (1989). A trichotomy method: Interactions of factors prolonging sequential and concurrent mental processes in stochastic PERT networks. Journal of Mathematical Psychology, 33, 328-347.

Shiffrin, R. M. (1970). Memory search. In D. A. Norman (Ed.), Models of human memory. New York: Academic Press.

Snodgrass, J. G., & Townsend, J. T. (1980). Comparing parallel and serial models: Theory and implementation. Journal of Experimental Psychology: Human Perception and Performance, 6, 330-354.

Sternberg, S. (1969). The discovery of processing stages: Extensions of Donders' method. Acta Psychologica, 30, 276-315.

Townsend, J. T. (1969). Mock parallel and serial models and experimental detection of these.  Purdue Centennial Symposium on Information Processing.  Purdue University: Purdue University Press.

Townsend, J. T. (1972). Some results concerning the identifiability of parallel and serial processes. British Journal of Mathematical and Statistical Psychology, 25, 168-199.

Townsend, J. T. (1974). Issues and models concerning the processing of a finite number of inputs. In B. H. Kantowitz (Ed.), Human Information Processing: Tutorials in Performance and Cognition (pp. 133-168). Hillsdale, NJ: Erlbaum Press.

Townsend, J. T. (1976a). Serial and within-stage independent parallel model equivalence on the minimum completion time. Journal of Mathematical Psychology, 14, 219-238.

Townsend, J. T. (1976b). A stochastic theory of matching processes. Journal of Mathematical Psychology, 14, 1-52.

Townsend, J. T. (1990). Serial vs. parallel processing: Sometimes they look like tweedledum and tweedledee but they can (and should) be distinguished. Psychological Science, 1, 46-54.

Townsend, J. T., & Ashby, F. G. (1983). The Stochastic Modeling of Elementary Psychological Processes. Cambridge: Cambridge University Press.

Townsend, J. T., & Evans, R. (1983). A systems approach to parallel-serial testability and visual feature processing. In H. G. Geissler (Ed.), Modern Issues in Perception (pp. 166-189). Berlin: VEB Deutscher Verlag der Wissenschaften.

Townsend, J. T. & Fific, M. (2004). Parallel & serial processing and individual differences in high-speed scanning in human memory. Perception & Psychophysics, 66, 953-962.

Townsend, J. T., & Nozawa, G. (1995). Spatio-temporal properties of elementary perception: An investigation of parallel, serial, and coactive theories. Journal of Mathematical Psychology, 39, 321-359.

Townsend, J. T. & Wenger, M.J. (2004). The serial-parallel dilemma: A case study in a linkage of theory and method. Psychonomic Bulletin & Review, 11, 391-418.

Townsend, J. T.,Yang, H.& Burns, D. M. (2011). Experimental discrimination of the world's simplest and most antipodal models: The parallel-serial issue. In Hans Colonius and Ehtibar Dzhafarov (Ed.s), Descriptive and Normative Approaches to Human Behavior in the Advanced Series on Mathematical Psychology. Singapore: World Scientific.

Vorberg, D., & Ulrich, R. (1987). Random search with unequal search rates: Serial and parallel generalizations of McGill's model. Journal of Mathematical Psychology, 31, 1-23.

Yang, H., Fific, M., & Townsend, J.T. (2013). Survivor interaction contrast wiggle predictions of parallel and serial models for an arbitrary number of processes. Journal of Mathematical Psychology, 58, 21-32.

Zhang, R., & Dzhafarov, E.N. (2015). Noncontextuality with marginal selectivity in reconstructing mental architectures. Frontiers in Psychology: Cognition 1:12 doi: 10.3389/fpsyg.2015.00735.